\documentclass[oldversion]{aa} 
\usepackage{natbib}
\usepackage{graphicx}
\usepackage{txfonts}
\usepackage{calc}
\usepackage{color}
\usepackage{rotating}
\usepackage[OT2, T1]{fontenc}
\usepackage[USenglish]{babel} 
\usepackage{pdflscape}

\begin{document}

\title{The dark side of solar photospheric G-band bright points}
\titlerunning{The dark side of bright points} 

\author{T.~L. Riethm\"uller\inst{1}
   \and S.~K. Solanki\inst{1,2}
   }

\institute{Max-Planck-Institut f\"ur Sonnensystemforschung (MPS),
           Justus-von-Liebig-Weg 3, 37077 G\"ottingen, Germany
     \and
           School of Space Research, Kyung Hee University,
           Yongin, Gyeonggi, 446-701, Republic of Korea\\
           \email{riethmueller@mps.mpg.de}
          }

\date{Received; accepted}

\abstract
{
Bright small-scale magnetic elements found mainly in intergranular lanes at the solar surface are named bright points (BPs).
They show high contrasts in Fraunhofer G-band observations and are described by nearly vertical slender flux tubes or sheets.
A recent comparison between BP observations in the ultraviolet (UV) and visible spectral range recorded with the balloon-borne
observatory \textsc{Sunrise} and state-of-the-art magnetohydrodynamical (MHD) simulations revealed a kiloGauss magnetic field
for 98\% of the synthetic BPs. Here we address the opposite question, namely which fraction of pixels hosting kiloGauss fields
coincides with an enhanced G-band brightness. We carried out 3D radiation MHD simulations for three magnetic activity levels
(corresponding to the quiet Sun, weak and strong plage) and performed a full spectral line synthesis in the G-band. Only 7\% of
the kiloGauss pixels in our quiet-Sun simulation coincide with a brightness lower than the mean quiet-Sun intensity, while 23\%
of the pixels in the weak-plage simulation and even 49\% in the strong-plage simulation are associated with a local darkening.
Dark strong-field regions are preferentially found in the cores of larger flux patches that are rare in the quiet Sun, but more
common in plage regions, often in the vertices of granulation cells. The significant brightness shortfall in the core of larger
flux patches coincide with a slight magnetic field weakening. KiloGauss elements in the quiet Sun are on average brighter than
similar features in plage regions. Almost all strong-field pixels display a more or less vertical magnetic field orientation.
Hence in the quiet Sun, G-band BPs correspond almost one-to-one with kiloGauss elements. In weak plage the correspondence is still
very good, but not perfect.
}

\keywords{Sun: magnetic fields --- Sun: photosphere --- Sun: magnetohydrodynamics (MHD)}

\maketitle

\section{Introduction}

Photospheric bright points (BPs) are small-scale brightness enhancements usually formed in intergranular lanes on the Sun's surface.
They were first observed by \citet{Dunn1973} in H$\alpha$ line wing images and later by \citet{Mehltretter1974} who, from a comparison between
Ca\,{\sc ii}~K images with a Kitt Peak magnetogram, concluded that BPs are magnetic structures. With sizes between 70\,km and 600\,km
\citep{Berger1995,SanchezAlmeida2004,Beck2007,Utz2009,Keys2013,Riethmueller2014} they belong to the smallest known magnetic features and often they are right at or
even below the resolution limit of modern solar telescopes. They typically live for a few 100\,s \citep{Muller1983,SanchezAlmeida2004,Keys2011,Jafarzadeh2013,Keys2014}.
The physical structure underlying BPs are magnetic elements, which in general have kiloGauss field strengths \citep[e.g.,][]{Stenflo1973,Rabin1992,Rueedi1992,Riethmueller2014}.

Over a solar activity cycle the contribution of the brightness excess of magnetic elements to solar irradiance is believed to overcompensate the darkening produced by
sunspots \citep{Krivova2003,Domingo2009,Froehlich2013,Solanki2013}. Hence the magnetic elements are thought to be responsible for an increased total solar irradiance (TSI)
during magnetic activity maxima of the Sun \citep{Willson1988}. The influence of TSI variations (on time scales of decades or more) on the terrestrial climate is under
debate \citep{London1994,Larkin2000,Gray2010,Haigh2010,Ermolli2013}. In particular the variations in the ultraviolet (UV) are much larger than for longer wavelengths
\citep{Krivova2006,Harder2009} and hence the UV properties of BPs are of great potential importance \citep{Riethmueller2010}. However, many of their properties are not
known with sufficient accuracy.

\citet{Criscuoli2014} studied G-band BPs in MHD simulations of the Stagger code \citep{Galsgaard1996} and determined a bimodal distribution of their
magnetic field strength. The first peak below 100\,G corresponds to bright granules misidentified as BPs and the second at about 1500\,G to magnetic
features. \citet{Utz2013} detected BPs in G-band observations recorded with the Solar Optical Telescope aboard the \textsc{Hinode} satellite
\citep{Tsuneta2008} and retrieved their field strength distribution from Milne-Eddington inversions of simultaneously acquired spectropolarimetric
data. They also found bimodal distributions of the field strength for the BPs, both in the quiet Sun and in an active region.

Similarly \citet{Riethmueller2014} reported that in an MHD simulation with an averaged field strength of 30\,G at the solar surface,
representative of the quiet Sun, 98\% of the synthetic BPs harbor kiloGauss field strengths. They also carefully compared a number of
spectropolarimetric parameters in the MHD simulation with those in high-resolution observations obtained by the IMaX instrument
\citep{MartinezPillet2011} on the \textsc{Sunrise} balloon-borne observatory \citep{Solanki2010,Barthol2011,Berkefeld2011,Gandorfer2011}.
This comparison showed that both can be reconciled if spatial and spectral degradation due to the instrument is meticulously taken into account.

Here we follow up on this work by considering the opposite question: Do all quiet-Sun kiloGauss elements lead to bright features or are there small
dark, strong-field features as well? We note that \citet{Riethmueller2014} concentrated on degraded MHD data (in order to allow direct comparisons
with the observations), while we now focus on undegraded MHD data that are noise-free and have high spatial resolution. We are able to do that
since the validity of the simulations was already tested versus seeing-free observations by \citet{Riethmueller2014}.

\section{Simulation and spectral synthesis}

We used the MURaM\footnote{The {\bf M}ax Planck Institute for Solar System Research / {\bf U}niversity of Chicago {\bf Ra}diation {\bf M}agneto-hydrodynamics code.} code
for realistic simulations of the radiative and magneto-hydrodynamical processes in the solar photosphere and the uppermost layers of the convection zone. MURaM is a
three-dimensional, non-ideal, compressible MHD code, which includes non-gray calculations of the radiative transfer of energy under the assumption of local thermal equilibrium
\citep{Voegler2005b}. At the bottom boundary of the simulation box a free in- and outflow of matter was allowed under the constraint of total mass conservation, while
at the upper boundary the vertical convective fluxes of mass, energy and horizontal momentum vanish, and the magnetic field lines are assumed to be vertical.
In the horizontal directions we used periodic boundary conditions. Our simulation box covers 6\,Mm\,$\times$\,6\,Mm in the horizontal directions, with a
cell size of 10.4\,km, while it covers 1.4\,Mm in the vertical direction, with a 14\,km cell size. On average, optical depth unity for the continuum at 500\,nm is
reached 500\,km below the upper boundary of the box.

Our initial condition was a statistically relaxed purely hydrodynamical simulation in which we introduced a unipolar homogeneous vertical magnetic field, $\langle B_{\rm{z}} \rangle$.
To reach a statistically stationary state again, we ran the simulation for a further 3 hours of solar time. Ten snapshots were then analyzed for each simulation run, one every
5 minutes of solar time, so that the snapshots can be considered to be nearly statistically independent. For this study we calculated ten snapshots taken from a simulation run
with an initial $\langle B_{\rm{z}} \rangle = 30$\,G (simulating a quiet-Sun region), another ten snapshots with $\langle B_{\rm{z}} \rangle = 200$\,G (simulating a weak plage
region), and finally ten snapshots  with $\langle B_{\rm{z}} \rangle = 400$\,G (strong plage). More details about the MHD simulation runs analyzed here can be found
in \citet{Riethmueller2014}.

The SPINOR\footnote{The {\bf S}tokes-{\bf P}rofiles-{\bf IN}version-{\bf O}-{\bf R}outines.} inversion code was used in its forward computation mode
\citep{Solanki1987,Frutiger2000a,Frutiger2000b} to compute synthetic Stokes spectra of the G-band, a spectral range around 430.5\,nm dominated by
lines of the CH molecule. This spectral range was selected since G-band brightenings have been used as proxies of small-scale magnetic features for a
long time and a wealth of observations are available \citep[e.g.,][]{Muller1984,Berger1995,Title1996,Wiehr2004,Zakharov2005,Beck2007,Berger2007,Bonet2012,Keys2014}.
We used a Lorentz profile having a full width at half maximum (FWHM) of 1\,nm as a typical filter transmission profile, multiplied it point by point with the
synthetic intensity profiles, and summed up the products. These scaler products gave the G-band intensity at a spatial pixel of a synthetic image.
Additional information about the spectral synthesis of the G-band can be found in \citet{Shelyag2004}, cf. \citet{Schuessler2003}. All computations
refer to the center of the solar disk.

\section{Results}

\subsection{Relationship between intensity and magnetic field strength}

\subsubsection{Full spatial resolution}
\begin{figure}
   \centering
   \includegraphics[width=\linewidth]{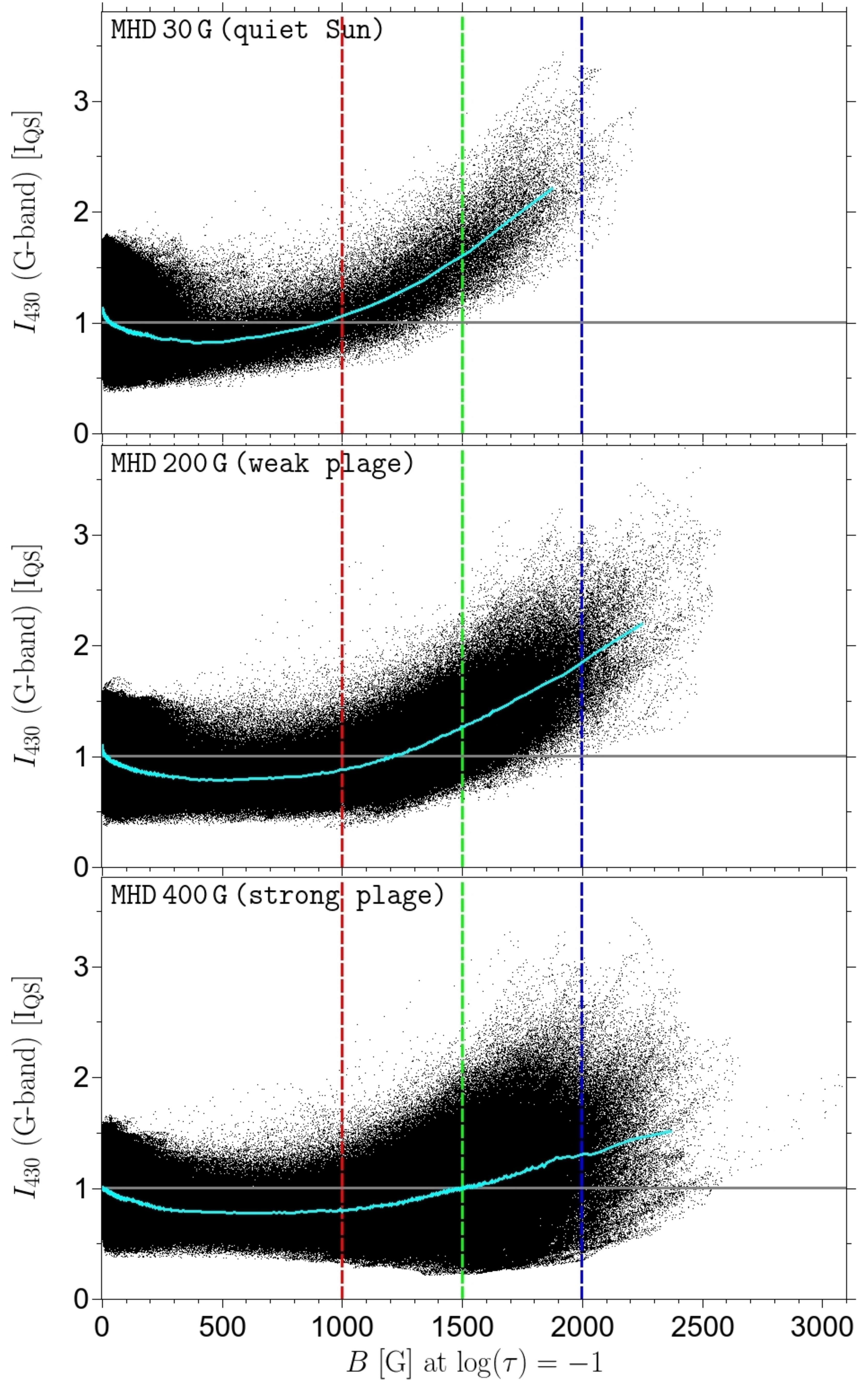}
   \caption{Scatter plots of the G-band intensity versus the magnetic field strength at $\log(\tau)=-1$ (black dots). The cyan lines connect
   binned values. The top, middle, and bottom panel corresponds to MHD simulations with an initial mean vertical flux density of 30\,G, 200\,G, and 400\,G, respectively.
   The horizontal line in each frame marks the mean quiet-Sun intensity and the vertical lines indicate field strengths of 1000\,G, 1500\,G, and 2000\,G.}
   \label{FigScatterPlots}
   \end{figure}

   Scatter plots of the G-band intensity versus the magnetic field strength are displayed in Fig.~\ref{FigScatterPlots} (black dots). All pixels of all
   snapshots contributed to this figure (576 pixels $\times$ 576 pixels $\times$ 10 snapshots). A binning of the scatter plots was applied by
   averaging groups of 5000 data points with similar $B$ (cyan lines in Fig.~\ref{FigScatterPlots}).
   The binned brightness drops below the mean quiet-Sun value, $I_{\rm{QS}}$ (taken to be the average intensity of the 30\,G simulation),
   for intermediate field strengths, before it rises to well above this value for kiloGauss fields. The scatter in intensity for the kiloGauss fields
   increases considerably with $\langle B_{\rm{z}} \rangle$. The magnetic field strength, used to produce the scatter plots, was taken at an optical depth of $\log(\tau)=-1$ because
   field strengths retrieved by Milne Eddington inversions of, e.g., the Fe\,{\sc i} 525.02\,nm line correspond to roughly this optical depth \citep{OrozcoSuarez2010}.

   This choice also allows a direct comparison of our 200\,G simulation (middle panel of Fig.~\ref{FigScatterPlots}) with the one carried out by \citet{Shelyag2004}
   (see the left panel of their Fig.~4). These authors calculated MURaM snapshots for an initial $\langle B_{\rm{z}} \rangle$ of 200\,G, but used
   a coarser computational grid with a cell size of 20.8\,km in both horizontal directions. The qualitative behavior of their scatter plot
   is very similar to ours. They found a maximum field strength of 2700\,G, which is also quite close to the value of 2600\,G
   that we retrieved from our 200\,G simulation. In contrast to these agreements, \citet{Shelyag2004} found a maximum G-band intensity value of
   2.2 times the mean quiet-Sun value, while we obtain a considerably higher value of $3.8\,I_{\rm{QS}}$. \citet{Shelyag2004} also considered a 10\,G simulation
   as quiet-Sun reference and found a maximum field strength of 1900\,G and a maximum G-band intensity of 1.6\,$I_{\rm{QS}}$, while we used an averaged field strength
   of 30\,G to simulate the quiet Sun and found 2300\,G and 3.4\,$I_{\rm{QS}}$ as the maximum values, but this time a direct comparison is difficult because both,
   the mean vertical field strength and the cell size are different between the simulations.
   
   A comparison between our quiet-Sun simulation (top panel of Fig.~\ref{FigScatterPlots}) and our plage simulations (middle and bottom panels of Fig.~\ref{FigScatterPlots})
   reveals not just that there are many more pixels with kiloGauss fields in the latter, but that these pixels are on average also somewhat less bright. Thus, the
   fraction of pixels possessing an intensity lower than $I_{\rm{QS}}$ (below the gray horizontal line) and a kiloGauss magnetic field (right of the red vertical line)
   increases from 0.13\% for the quiet-Sun simulation to 4.2\% for the weak-plage simulation and to 15.8\% for the strong-plage simulation. For a magnetic field threshold
   of 1500\,G (green vertical line) we find no dark pixels anymore in the quiet-Sun snapshots but still 0.24\% (4.6\%) of all pixels of the weak- (strong-) plage simulation.
   
   To get a better insight into this effect we calculated G-band intensity histograms
   of only those pixels whose magnetic field strength exceeds a certain threshold, $B_T$. These histograms, $H_{B_T}(I_{430})$, are then integrated over the intensity
   \begin{equation}\label{Eq_HistIntegral}
   F_{B_T}(I_{430})=\frac{1}{N_{B_T}} \int\limits_{I_{430}}^{\infty} H_{B_T}(I_{430}^\prime)\,dI_{430}^\prime ,
   \end{equation}
   where $1/N_{B_T}$ implies normalization to the total number of pixels that harbor a field strength of at least $B_T$,
   \begin{equation}\label{Eq_HistIntegralNorm}
   N_{B_T}=\int\limits_{0}^{\infty} H_{B_T}(I_{430}^\prime)\,dI_{430}^\prime ,
   \end{equation}
   so that $F_{B_T}(I_{430})$ gives the fraction of strong-field pixels having a G-band intensity of at least $I_{430}$.
   
   \begin{figure}
   \centering
   \includegraphics[width=\linewidth]{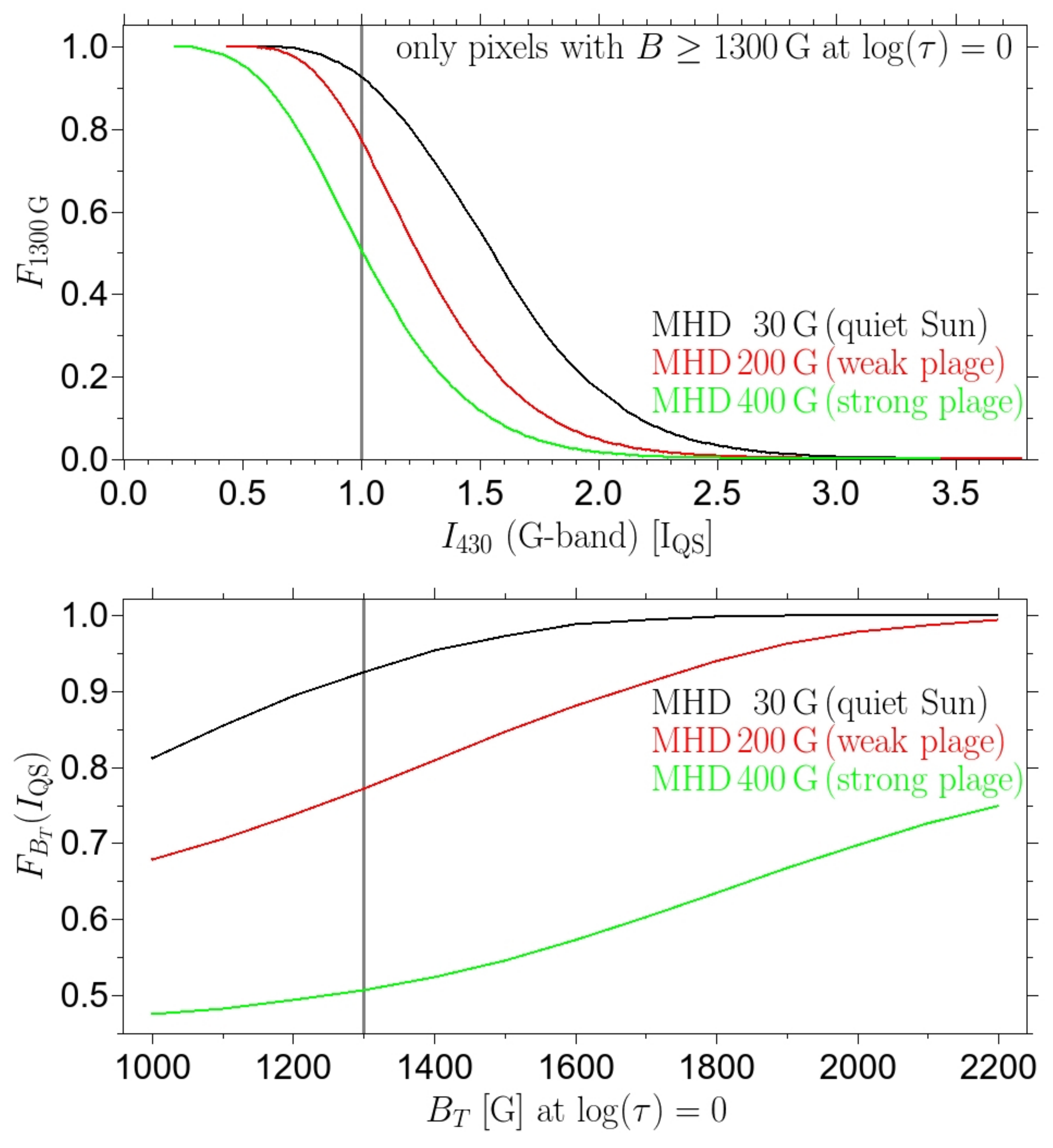}
   \caption{Top panel: Integrated G-band intensity histograms, $F_{B_T}$, of strong-field pixels with $B\ge1300$\,G (see main text for definition).
   The mean quiet-Sun intensity is indicated by the vertical line. Bottom panel: Variation of $F_{B_T}(I_{\rm{QS}})$ with $B_T$, where $I_{\rm{QS}}$ is the
   mean quiet-Sun intensity and $B_T$ is the magnetic field threshold. $F_{B_T}(I_{\rm{QS}})$ gives the fraction of all pixels having a field strength
   greater than $B_T$, which also have an intensity greater than $I_{\rm{QS}}$. The vertical line indicates the lowest $B_T$ value at optical
   depth unity that corresponds to a kiloGauss field at the height of average line formation of Fe\,{\sc i} 525.02\,nm.
   The black, red, and green lines refer to the snapshots taken from the 30\,G, 200\,G, and 400\,G simulation.}
   \label{FigHistIntegral}
   \end{figure}

   Since the gas pressure decreases with height in the solar photosphere, flux tubes expand with height and form a canopy.
   This is a region where a vertical ray coming from the top passes through the magnetized atmosphere of the expanding flux tube in the
   upper photosphere, then hits the nearly field-free atmosphere below the canopy before penetrating the $\tau = 1$ surface
   and entering the solar interior \citep[e.g.,][]{Solanki1989}. Such rays often show a kiloGauss field combined with a low brightness due to the location of
   magnetic elements in intergranular lanes. By limiting our analysis to pixels harboring kiloGauss field strengths at the
   $\tau = 1$ surface we excluded pixels in the canopy. A scatter plot of the magnetic field strength at $\log(\tau)=0$ versus
   the field strength at $\log(\tau)=-1$ taken from our quiet-Sun simulation (not shown) revealed that on average a field strength
   at optical depth unity of at least 1300\,G is needed to reach kiloGauss field strengths at $\log(\tau)=-1$, i.e. at the
   optical depth to which the field strengths retrieved by Milne Eddington inversions of the Fe\,{\sc i} 525.02\,nm line refer to.
   
   The top panel of Fig.~\ref{FigHistIntegral} displays $F_{B_T}(I_{430})$ for a magnetic field threshold at optical depth unity
   of $B_T=1300$\,G. These strong-field pixels of the 30\,G, 200\,G, and 400\,G simulation cover an intensity range of
   $0.54\,I_{\rm{QS}} - 3.44\,I_{\rm{QS}}$, $0.43\,I_{\rm{QS}} - 3.78\,I_{\rm{QS}}$, and $0.22\,I_{\rm{QS}} - 3.45\,I_{\rm{QS}}$, respectively.
   93\% of the kiloGauss pixels of the 30\,G snapshots reach an intensity higher than the mean quiet Sun, but only 77\% of
   the kiloGauss pixels of the 200\,G snapshots and just 51\% of the 400\,G simulation. We note that a fraction of 93\% in the quiet-Sun
   simulation means that 7\% of the kiloGauss pixels are darker than the mean quiet Sun. This fraction is equivalent to 0.06\% of all pixels
   (including pixels of any field strength) and hence lower than the 0.13\% mentioned above because we here exclude the dark canopy pixels.
   We also plotted $F_{1000\,\mathrm{G}}(I_{430})$, $F_{1500\,\mathrm{G}}(I_{430})$, and $F_{2000\,\mathrm{G}}(I_{430})$ (not shown) and
   found that the upper limit of the intensity ranges is not influenced by $B_T$ because the brightest pixels always possess very strong fields.
   The lower limit of the intensity ranges as well as the difference between the curves for the three simulations increases with $B_T$.
   
   In the bottom panel of Fig.~\ref{FigHistIntegral} we show the variation of the $F_{B_T}$ value with $B_T$ at the constant mean
   quiet-Sun intensity, $I_{430}=I_{\rm{QS}}$, i.e. it shows the fraction of pixels with $B>B_T$ that have $I>I_{\rm{QS}}$.
   Generally, the fraction of strong-field pixels reaching intensities higher than $I_{\rm{QS}}$ is larger in our quiet-Sun
   simulation than in our plage simulations. For the lower part of the displayed $B_T$ range the $F_{B_T}$ curves increase
   roughly linearly, while in the kiloGauss range the curves approach asymptotically the value 1.
   The smoothness of the three curves shows that the fraction of strong-field pixels having over-average brightnesses that we
   retrieved from our analysis depends smoothly on the chosen $B_T$ threshold.

\subsubsection{Reduced spatial resolution}
\begin{figure}
   \centering
   \includegraphics[width=\linewidth]{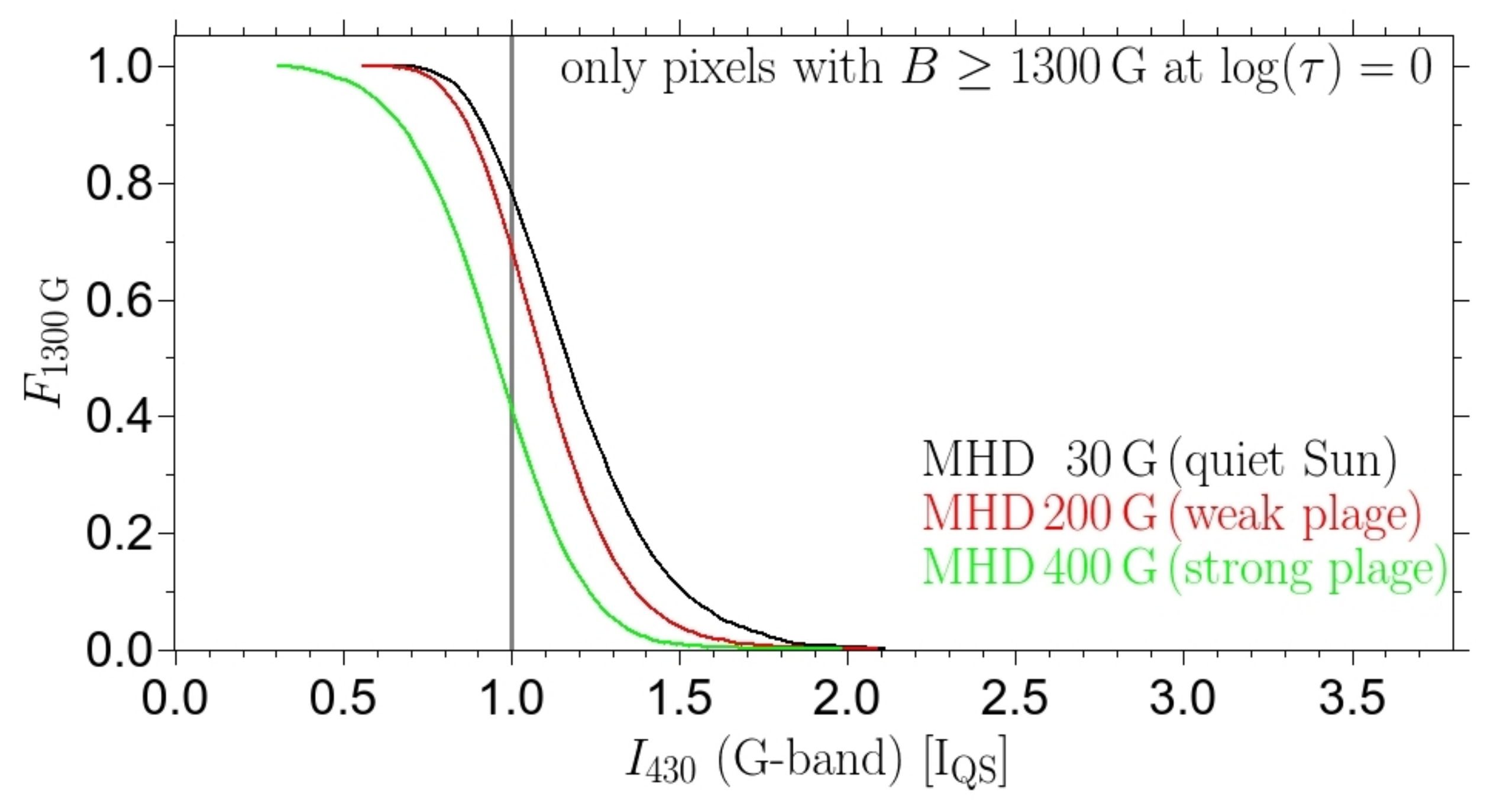}
   \caption{Same as the top panel of Fig.~\ref{FigHistIntegral} but for G-band data degraded to the diffraction limit of a 0.5\,m telescope.}
   \label{FigDegraded}
   \end{figure}

   To demonstrate the influence of the spatial resolution on the fraction of strong-field pixels in the form of G-band BPs, we convolved the
   G-band images with an Airy function that corresponds to the G-band wavelength, 430.5\,nm, and to a circular telescope aperture of
   diameter $D=0.5$\,m (e.g. \textsc{Hinode}/SOT), $D=1$\,m (\textsc{Sunrise}, SST), $D=1.5$\,m (\textsc{Gregor}, NST), and $D=4$\,m (DKIST), respectively.
   As an example, we plot the integrated histograms of the $D=0.5$\,m case in Fig.~\ref{FigDegraded}. Qualitatively, the curves for the three
   magnetic types of simulation show a behavior similiar to the undegraded case displayed in the top panel of Fig.~\ref{FigHistIntegral}.
   The degradation squeezes the curves towards lower intensities, so that the fractions of strong-field pixels having $I>I_{\rm{QS}}$
   become smaller (intersections with the vertical line).
   
   The dependence of the fractions on the level of degradation is given in Table~\ref{FractionTable} and shows that the diffraction limit
   of a 4\,m telescope hardly affects the fraction of kiloGauss pixels, while the reduction for a 1\,m or an even smaller telescope is significant.
   We note that the cell size of our simulations corresponds roughly to half the diffraction limit of a 4\,m telescope in the G-band spectral range,
   so that a significant influence of a 4\,m telescope diffraction on the fraction of kiloGauss pixels in a simulation with a cell size smaller than
   10\,km cannot be ruled out.

   \begin{table}
   \caption{Fraction of kiloGauss field in the form of G-band bright points for various telescope sizes.}
   \label{FractionTable}                                            
   \begin{tabular}{l l l l}                                         
   \hline                                                           
   \noalign{\smallskip}
   Degradation      & 30\,G       & 200\,G       & 400\,G         \\
                    & (quiet Sun) & (weak plage) & (strong plage) \\
   \hline                                                   
   \noalign{\smallskip}                                     
   undegraded       & 0.93        & 0.77         & 0.51           \\ 
   4.0\,m telescope & 0.92        & 0.77         & 0.50           \\ 
   1.5\,m telescope & 0.91        & 0.76         & 0.48           \\ 
   1.0\,m telescope & 0.88        & 0.74         & 0.47           \\ 
   0.5\,m telescope & 0.78        & 0.68         & 0.41           \\ 
   \hline                                                            
   \end{tabular}
   \end{table}
   
\begin{figure*}
   \centering
   \includegraphics[width=\linewidth]{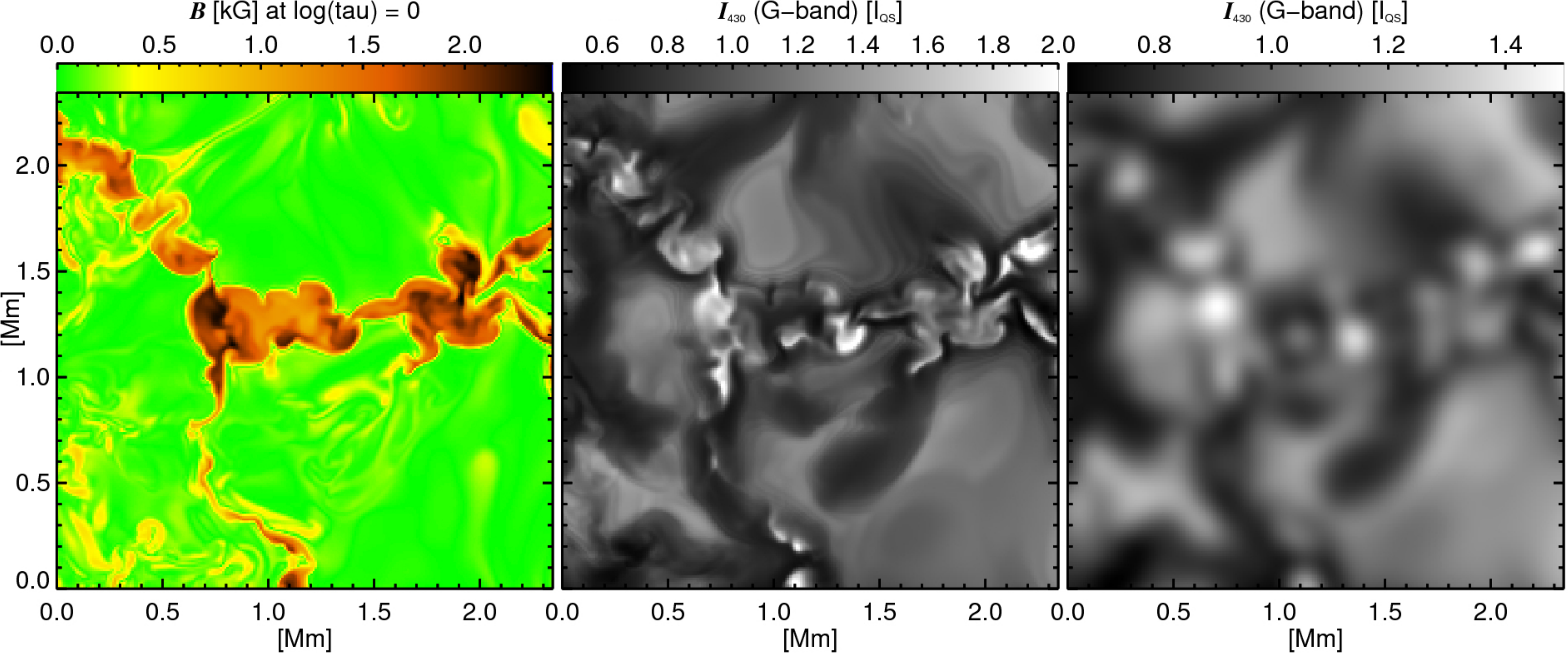}
   \caption{Magnetic field strength at optical depth unity (left panel), G-band intensity at full spatial resolution (middle panel)
   and as seen by an ideal 0.5\,m telescope (right panel) of a magnetic element of the 200\,G simulation harboring bright and dark strong-field pixels.}
   \label{FigDegBP}
   \end{figure*}

   Fig.~\ref{FigDegBP} displays the influence of the diffration at the aperture of a 0.5\,m telescope on a G-band intensity image
   of a kiloGauss element extended from 0.6 to 1.5\,Mm in x direction and from 1.1 to 1.4\,Mm in y direction (see the left panel of Fig.~\ref{FigDegBP}).
   The undegraded G-band intensity is mainly below the quiet-Sun average in the central part of the magnetic feature, while being mostly above the quiet-Sun average
   in the periphery of the feature, in particular at its left as well as its bottom-right edge. Nonetheless, at full spatial resolution
   the magnetic feature can be recognized as a single entity (see the middle panel of Fig.~\ref{FigDegBP}). This changes when the feature is
   observed with a 0.5\,m telescope (see the right panel of Fig.~\ref{FigDegBP}). At the reduced spatial resolution the feature appears as three
   isolated small BPs around the positions (0.7\,Mm,1.3\,Mm), (1.1\,Mm,1.2\,Mm), and (1.4\,Mm,1.2\,Mm), while the dark central part of the feature can
   no longer be recognized as belonging to the entity.

\subsection{Dark strong-field regions}

\subsubsection{Locations within the convection pattern}
\begin{figure}
   \centering
   \includegraphics[width=\linewidth]{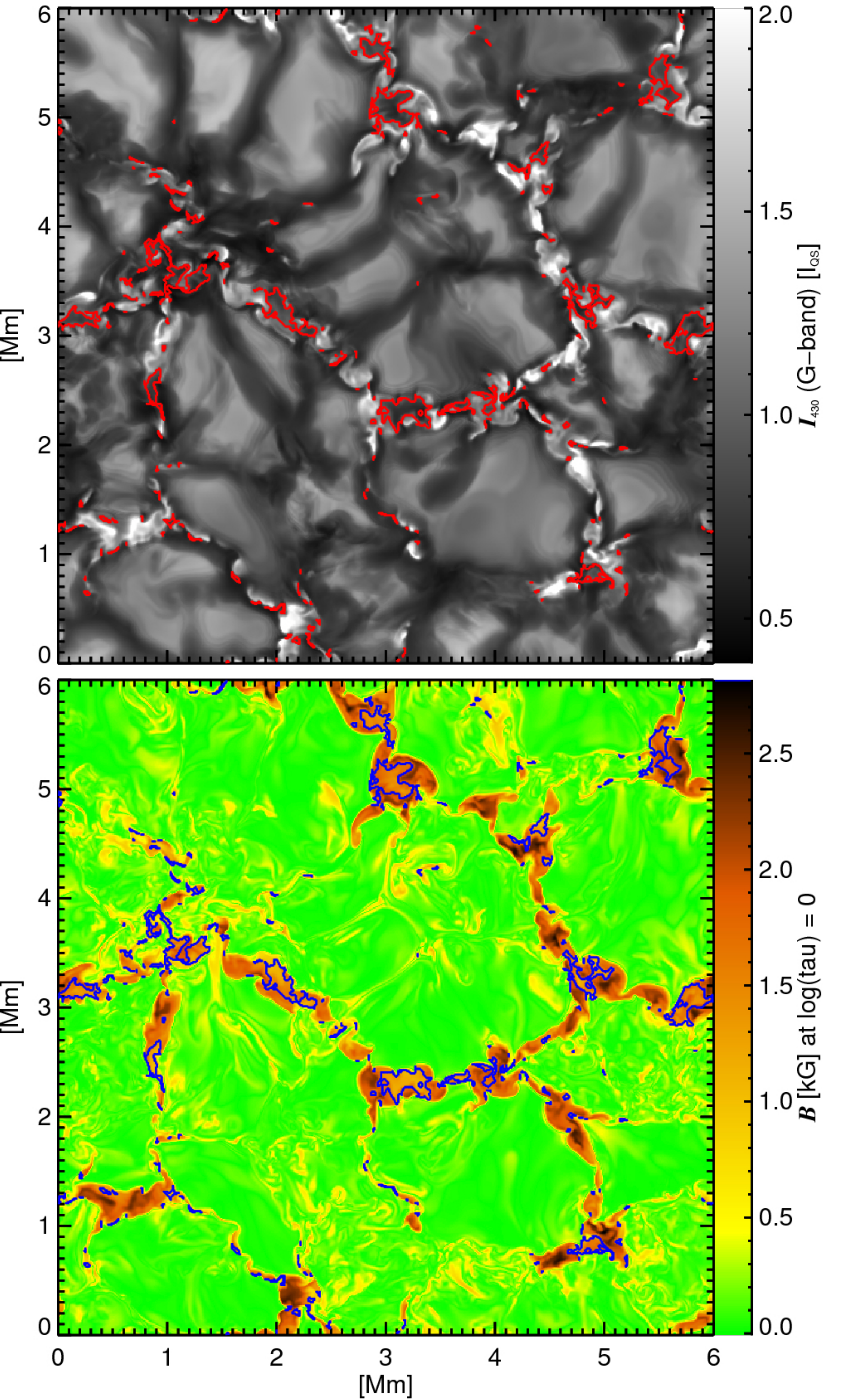}
   \caption{G-band intensity (top panel) and magnetic field strength at optical depth unity (bottom panel) of a 200\,G MHD simulation.
   The contour lines (red in the top panel and blue in the bottom panel) indicate dark, strong-field regions, i.e. $I_{\rm{430}} < I_{\rm{QS}}$ and $B_{\rm{\tau=1}} > 1000$\,G.
   The intensities of the G-band image are saturated at $2\,I_{\rm{QS}}$ for a better visibility of the granulation pattern.}
   \label{FigDarkStrongField}
   \end{figure}

   Figures~\ref{FigScatterPlots} and ~\ref{FigHistIntegral} reveal that even if the majority of kiloGauss pixels are brighter
   than the mean quiet Sun, a non-negligible fraction of them is dark. Here we analyze where those dark pixels are located within
   the solar granulation pattern. In Fig.~\ref{FigDarkStrongField} we show a G-band intensity map (top panel) and a map of the
   magnetic field strength at optical depth unity (bottom panel) of one of the weak-plage, 200\,G, snapshots. In agreement with earlier
   studies, these images show that strong-field regions are predominantly located in intergranular lanes, in particular the largest
   ones are often found at the vertices of three or more granulation cells. We over-plotted contour lines that indicate regions darker than
   the mean quiet Sun where the field strength is at least 1\,kG. These regions are mainly located in the cores of larger flux
   patches with horizontal sizes of $200-500$\,km (at their narrowest points), see, e.g., the patches at $(x,y)=(3.0\,\rm{Mm},\,5.1\,\rm{Mm})$ and $(3.2\,\rm{Mm},\,2.3\,\rm{Mm})$.
   In addition to these larger dark patches there are also smaller dark strong-field regions. These are preferentially found at the edges
   of flux patches, most probably because the G-band intensity is mainly formed somewhat higher than optical depth unity so that an inclination of
   the magnetic field can lead to a slight mismatch between the magnetic and the brightness structure.
   
\subsubsection{Magnetic field weakening}
\begin{figure*}
   \centering
   \includegraphics[width=\linewidth]{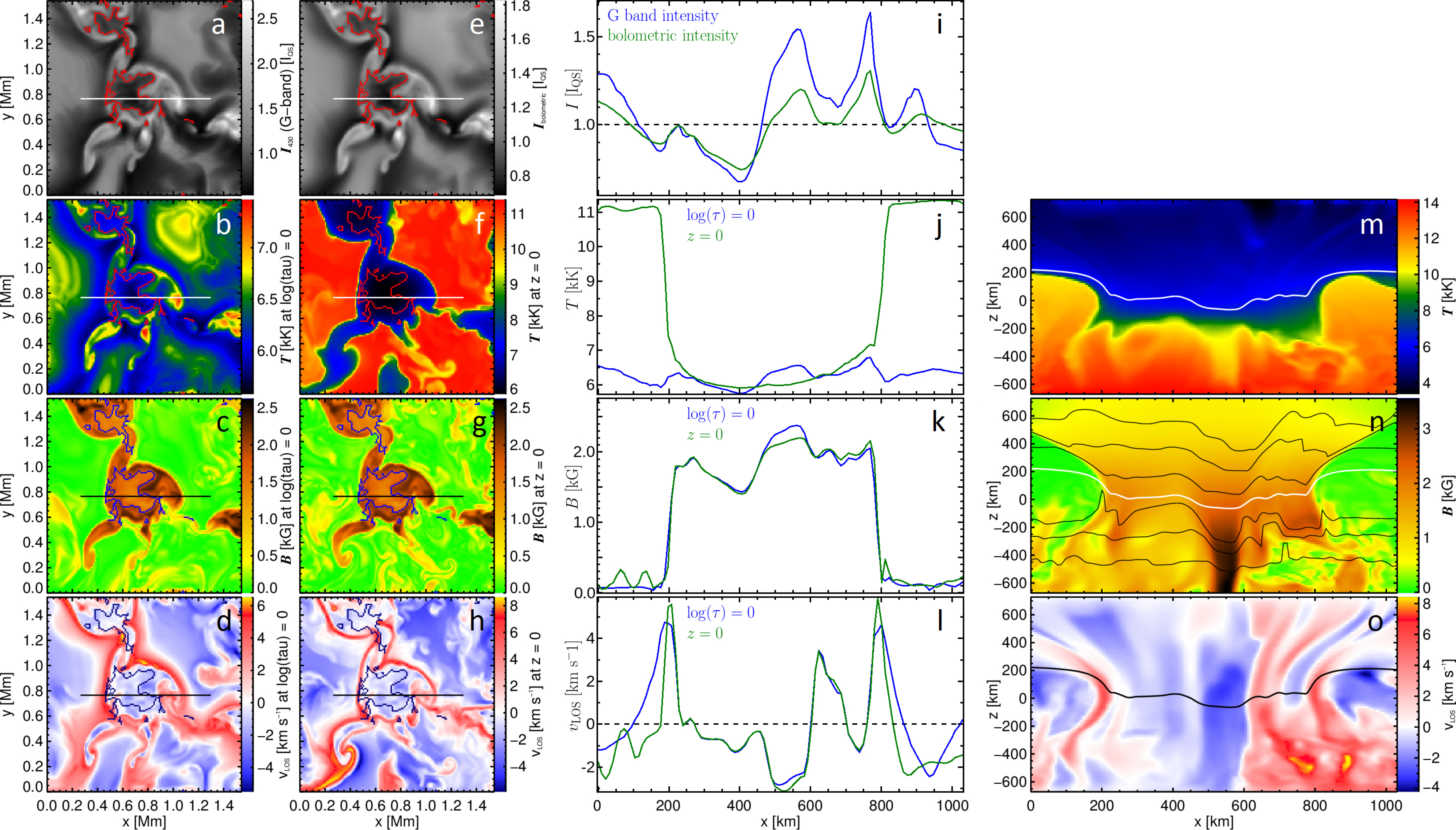}
   \caption{Maps of the G-band intensity (panel a), bolomeric intensity (panel e), temperature, magnetic field strength, and line-of-sight velocity at optical
   depth unity (panels b-d) and at constant geometrical height zero (panels f-h) of the flux patch at $(x,y)=(3.0\,\rm{Mm},\,5.1\,\rm{Mm})$ in Fig.~\ref{FigDarkStrongField}.
   Contour lines are the same as in Fig.~\ref{FigDarkStrongField}. The horizontal lines mark the position of the profiles plotted in panels i-l (blue lines for the
   first column of panels a-d and green lines for the second column of panels e-h) and the vertical cuts displayed in panels m-o. The optical depth unity level is
   marked by a white line in panels m and n and as a black line in panel o. The black lines in panel n are iso-density lines. Negative velocities correspond to upflows.}
   \label{FigVCut}
   \end{figure*}

   Often not only the intensity in the interior of large flux patches drops, but surprisingly also a slight magnetic field weakening
   can be found at these places (see the bottom panel of Fig.~\ref{FigDarkStrongField}). More insight into such magnetic field weakening
   in the interiors of large flux patches can be gained from Fig.~\ref{FigVCut}, which shows, among other things, vertical cuts
   through the flux patch at $(x,y)=(3.0\,\rm{Mm},\,5.1\,\rm{Mm})$ in Fig.~\ref{FigDarkStrongField} along the horizontal
   line shown in panels a-h of Fig.~\ref{FigVCut}. Both the G-band (panel a) and bolometric intensity (panel e) exhibit a clearly reduced brightness
   in the interior of the flux patch. The two intensity profiles (panel i) display a similar behavior, but the intensity contrast is considerably higher in
   the G-band, the main reason why G-band observations are frequently used. The temperature profile at optical depth unity (blue line in panel j) more or
   less follows the intensity profiles, while the temperature profile at zero geometrical height (green line in panel j) displays a significant depression
   that coincides with an enhanced magnetic field strength (panel k). The optical depth unity surface (e.g., white line in panel m) exhibits a clear
   Wilson depression, as expected for strong-field regions. The magnetic field weakening in the magnetic feature's interior is not only visible in the map
   at constant optical depth (panel c) but also in the map at constant geometrical height (panel g).
   
   The vertical cut in panel n conspicuously shows the flux-tube expansion above the $\tau=1$ line. The vertical cut also reveals that the magnetic
   field strength is rather inhomogeneously distributed across the flux tube. Compared with the strongest magnetic fields of around 2300\,G at
   $x=475$~to~$575\,\rm{km}$ the field strength at $x=300$~to~$400\,\rm{km}$ is only around 1400\,G. The weaker field is associated with a lower
   magnetic pressure which is approximately balanced by a higher gas pressure. This in turn is accompanied by a higher density and hence a raised
   $\tau=1$ level. Accordingly, the $\tau=1$ line is located at lower temperatures compared with the adjacent region and hence the region is darker.
   The same mechanism appears to be acting also in the other bigger dark patches.
   
   Horizontal maps of the line-of-sight (LOS) velocity (panels d and h of Fig.~\ref{FigVCut}) display downflow lanes surrounding the dark strong-field regions.
   These downflow lanes are narrower in the map at constant geometrical height. They are narrow but somewhat inclined with height, so that they appear broader
   at constant optical depth because of the sampling of a range of heights. While the majority of pixels in the dark strong-field region shown in the
   centers of the panels of Fig.~\ref{FigVCut} displays upflows, this in not always the case. Other dark strong-field regions usually harbour downflows or both, up- and downflows.
   The vertical cut in panel o exhibits a quite inhomogeneous velocity distribution with both upflows and downflows inside the flux patch. The strongest
   downflows are located at the edges of the strong-field magnetic feature and are the downflow lanes in which the feature is embedded. The strongest
   upflows coincide with the strongest magnetic field, although the weaker, darker field is also associated with an upflow. The transition between strong
   upflow and downflow regions does not lead to any distinctive feature in intensity.
   
\subsection{Relation between magnetic field strength and inclination}

\begin{figure}
   \centering
   \includegraphics[width=\linewidth]{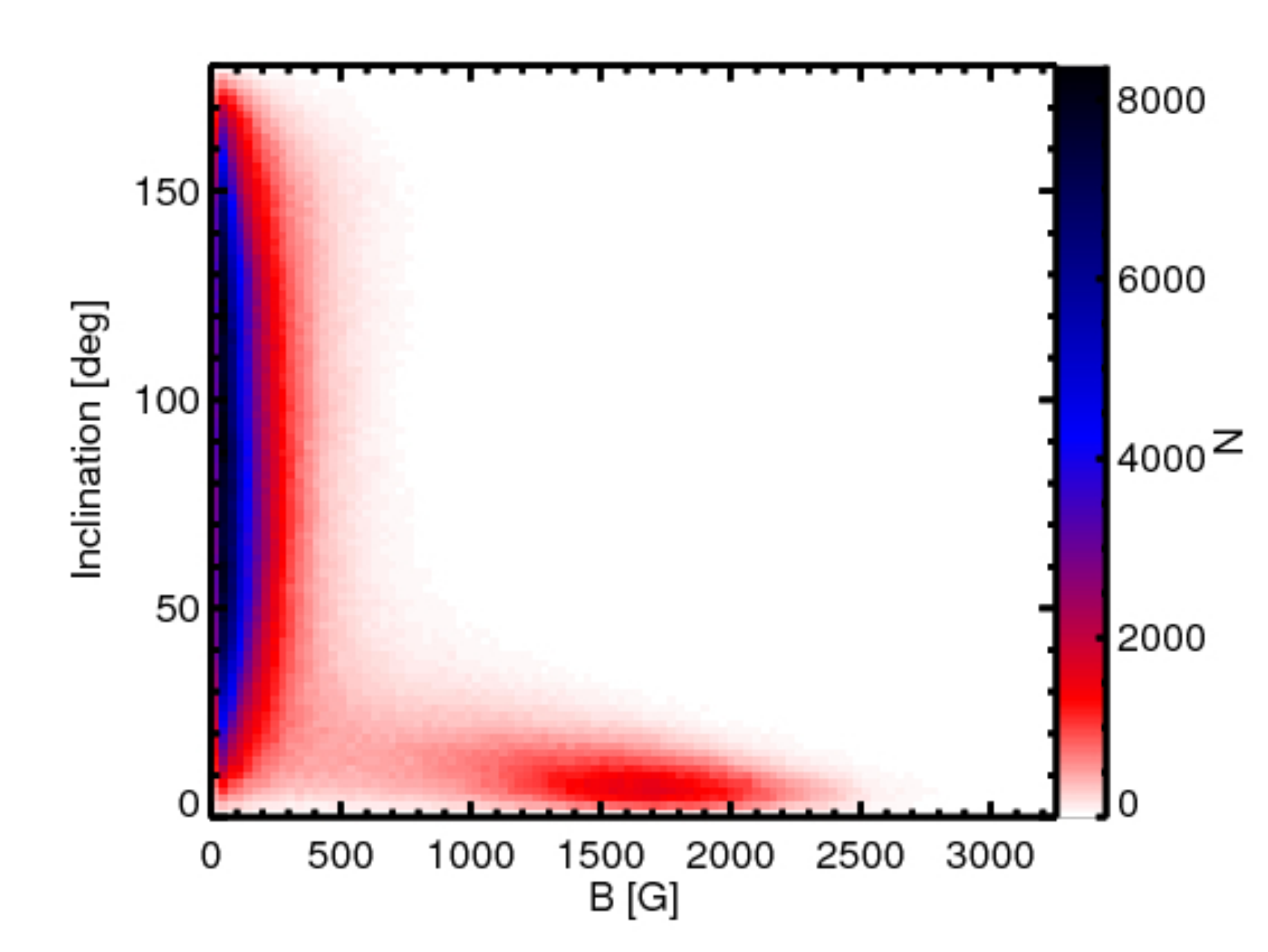}
   \caption{Two-dimensional histogram of the magnetic field strength and inclination of the weak-plage simulation at optical depth unity.}
   \label{FigHist2D}
   \end{figure}

   Finally, we calculated a two-dimensional histogram of the magnetic field strength and field inclination of all pixels of the weak-plage simulation at optical
   depth unity (Fig.~\ref{FigHist2D}). We found a bimodal distribution whose first population peaks at (30\,G, $86^{\circ}$), i.e. for almost horizontally oriented,
   very weak fields, while the second populaton has its peak at (1700\,G, $5^{\circ}$), i.e. for almost vertical strong fields. While the strong-field population displays
   a relatively narrow range of field inclinations (roughly between $0^{\circ}$-$20^{\circ}$), the angular distribution becomes asymptotically isotropic towards zero field strength.
   We also calculated such two-dimensional histograms for the optical depths $\log(\tau)=-1$ and $-2$ (not shown). While the position of the weak-field peak does not
   change much with optical depth, the position of the strong-field peak depends considerably on the optical depth, being at (1400\,G, $7^{\circ}$) and (880\,G, $13^{\circ}$) for
   $\log(\tau)=-1$ and $-2$, respectively. We note that the magnetic field distributions of our quiet-Sun simulation (not shown) also have a bimodal character, but
   because of the fewer strong-field pixels the strong-field population is much less pronounced.

\section{Summary and discussion}

   It has recently been shown that basically all BPs are associated with kilogass fields \citep{Riethmueller2014}. However, it was up to now not
   so clear which fraction of kiloGauss fields manifests itself in the form of BPs. With the aim of addressing this open question we simulated
   the upper convection zone and photosphere of the Sun by using the non-gray version of the MURaM code. We calculated ten snapshots each for a 
   mean vertical flux density of 30\,G (simulating the quiet Sun), 200\,G (weak plage), and 400\,G (strong plage), respectively. A full spectral
   line synthesis in the G-band was then carried out since this spectral range has been frequently used for observations. A typical filter profile
   was applied to the Stokes~$I$ profiles and G-band intensity images were retrieved.
   
   Binned scatter plots of the G-band intensity versus the magnetic field strength (see Fig.~\ref{FigScatterPlots}) displayed a monotonic relation
   for field strengths higher than about 500\,G. The analyses of \citet{Roehrbein2011} and \citet{Kahil2017} showed that the non-monotonic
   relation with a maximum at intermediate field strength found in scatter plots of observations at continuum wavelengths
   \citep[e.g.,][]{Lawrence1993,Topka1997,Kobel2011,Schnerr2011} is caused by image smearing due to the limited spatial
   resolution of observations. Our scatter plot from the weak-plage simulation revealed a similar range in field strength and a considerably larger range
   in intensity than the simulation carried out by \citet{Shelyag2004}, who used an almost identical setup, but with only half of the spatial resolution in the
   horizontal directions. Horizontal pressure balance leads to an evacuation of strong-field magnetic features and hence to a depression
   of the optical depth surface. In the theoretical picture of flux tubes used to describe such magnetic features the lateral inflow of radiation through
   the walls of this depression makes the flux tube hot and bright \citep{Spruit1976,Deinzer1984} with the G-band being particularly sensitive to such
   brightenings due to the preponderance of CH lines \citep{SanchezAlmeida2001,Steiner2001}. A doubling in spatial resolution increases in
   particular the upper limit of the intensity range significantly, which we believe is due to the thinner tube walls produced by the higher resolution
   which in turn leads to a more effective heating mechanism.
   
   Histograms of the G-band intensity were calculated for pixels exceeding a certain threshold in magnetic field strength. By taking the field strength
   at the $\tau=1$ surface we excluded pixels in the canopy of magnetic elements. We chose a field strength threshold of 1300\,G at $\tau=1$, which corresponds to mainly
   kiloGauss pixels at $\tau=0.1$, the optical depth to which field strengths retrieved by Milne Eddington inversions roughly refer to. These histograms
   were then integrated over the intensity starting from a given intensity threshold. This gave us the fraction of strong-field pixels that reached a
   certain minimum brightness (see Fig.~\ref{FigHistIntegral}). A comparison between such integrated histograms from our simulations of different mean flux densities
   revealed that kiloGauss magnetic features are on average less bright in plage regions than in the quiet Sun. This is consistent with observations
   \citep{Solanki1984,Solanki1986,Lawrence1993,Ishikawa2007,Kobel2011,Romano2012,Ji2016} and suggests that at least part of the lower brightness of simulation
   snapshots having larger magnetic flux \citep{Voegler2005a} is due to the less bright magnetic features, although a part may be due to the suppression of convection
   by the magnetic field \citep[e.g.,][]{Ishikawa2007,Kobel2012,Criscuoli2013}.
   
   We find that in the quiet Sun the G-band BPs are a very good guide to kiloGauss magnetic fields in that roughly 93\% of the area
   covered by kiloGauss magnetic fields is brighter than average in the G-band (in the absence of spatial smearing and scattered light).
   This also means that 93\% of the kiloGauss flux is captured by G-band BPs. This fraction decreases with increasing magnetic flux,
   being about 77\% for the 200\,G weak-plage simulation and just 51\% for the 400\,G strong-plage simulation. Spatial smearing
   also decreases the fraction, e.g., for an ideal 0.5\,m telescope it goes down to only 78\%, 68\%, and 41\% for the 30\,G, 200\,G, and
   400\,G simulation, hence the conclusion of \citet{Berger1995} from their active region observation with a 0.5\,m telescope that less
   than 50\% of the small-scale magnetic flux can be identified by visible proxies such as BPs or pores is not in contradiction to
   the resuls of this study, but can be well explained by the increased mean magnetic flux in the active region and by the limited spatial
   resolution.

   The smaller fraction of bright kiloGauss features in simulations with highest $\langle B_{\rm{z}} \rangle$ has to do with the on average
   larger magnetic features in such areas. These features are often found to have a slightly dark core in our simulations.
   The lateral inflow of heat through the walls of a flux tube is balanced by the tube's radiative losses at the solar surface.
   With increasing diameter of a flux patch, $d$, the wall area grows with $d$ while the area of radiative losses grows with $d^2$.
   Thus the heating due to the lateral inflow of radiation becomes ineffective for larger flux tubes so that their core becomes dark,
   while their peripheral regions remain bright (see top panel of Fig.~\ref{FigDarkStrongField}). The features we have found to
   have slightly dark cores are smaller and less dark than pores. They correspond more closely to 'knots' and 'points' features
   found in the older literature \citep{Beckers1968,Knoelker1988}, or to 'flowers' and 'micropores' in more recent publications
   \citep{Rouppe2005,Narayan2010}.
   
   During the last few years there has been a debate about the balance between the horizontal and the vertical magnetic flux in the quiet Sun. While some authors find
   a clear preference for horizontal magnetic fluxes in their observations \citep[e.g.,][]{OrozcoSuarez2007,Lites2008}, others claim an isotropic angular distribution
   of the magnetic field \citep[e.g.,][]{MartinezGonzalez2008,AsensioRamos2009}, still others find a broad range of field inclinations with a preference of vertical
   orientations \citep[e.g.,][for a review]{Khomenko2003,Beck2009,Borrero2015}. The fact that the magnetic structuring continues on scales well below the spatial resolution of modern solar
   telescopes is seen by \citet{Stenflo2010} as the reason for these contradictory results. \citet{Stenflo2010} used quiet-Sun data obtained from \textsc{Hinode}/SP
   to calculate a scatter plot of the blue lobe Stokes~$V$ amplitude derived from the Fe\,{\sc i} 630.15\,nm line versus the corresponding Stokes~$V$ amplitude for the
   Fe\,{\sc i} 630.25\,nm line and finds two different magnetic populations.
   
   While the interpretation of the two populations identified by \citet{Stenflo2010} is difficult (in terms of deriving the two relevant physical quantities,
   magnetic field strength and inclination, from Stokes signals) and still under debate \citep[see][]{Steiner2012}, we also obtained two different magnetic populations
   from our analysis, but with a much more straight-forward interpretation, because our MHD data set gives direct access to the relevant physical quantities,
   is free of noise, and has a much higher spatial resolution. The two-dimensional histograms that we calculated from the magnetic field strength and inclination
   of our weak-plage simulation shows one population formed exclusively of weak-field pixels displaying a broad range of inclinations. It also
   shows another population of pixels with mainly strong and more or less vertical fields. A look at the bottom panel of Fig.~\ref{FigDarkStrongField}
   reveals that all the strong-field pixels lie in the intergranular lanes. The distribution of the inclinations of the weak-field population
   converges to an isotropic angular distribution at very small flux densities, which was also obtained by \citet{Lagg2016} from their MHD simulations and by
   \citet{Stenflo2010} from the linear-to-circular polarization ratio of the \textsc{Hinode}/SP data he employed.
      
   The peak field strength of the strong-field population of our weak-plage simulation decreases and the peak inclination weakly increases with atmospheric height,
   which supports the picture of a thin flux tube expanding with height. From spatially coupled Stokes inversions of \textsc{Hinode}/SP data recorded in a plage region,
   \citet{Buehler2015} obtained at $\log(\tau)=-0.9$ a typical field strength of 1520\,G at inclination angles of $10^{\circ}$-$15^{\circ}$. This agrees well with our results
   retrieved from the weak-plage simulation at $\log(\tau)=-1$, where we determined the peak of the kiloGauss population to be located at 1400\,G and $7^{\circ}$.
    
   \citet{Criscuoli2014} used 3D MHD snapshots (calculated with the Stagger code) having a mean magnetic flux density of
   200\,G and a 22\,km cell size in the horizontal direction. They found a bimodal magnetic field distribution for small-scale
   bright features at an optical depth of $\log(\tau)=-1$. The first peak lay at a field strength close to zero and was caused by bright granules
   misidentified as BPs and the second peak lay at around 1500\,G, which is close to our strong-field peak at that depth, located at 1400\,G.
   Since we used a different MHD code and a spatial resolution twice as high we conclude that this result seems to be relatively insensitive to
   the details of the simulation.
   
   We mainly considered the question which fraction of the kiloGauss-field regions can be captured by BPs, depending on the spatial
   resolution and magnetic activity of the observed scene. While in this study we answered this question purely by analyzing MHD simulations, a future
   study should compare the properties of BPs between simulations and high-resolution active-Sun observations, similar to the study of
   \citet{Riethmueller2014} for the quiet Sun. A good opportunity might be a combination of active-region observations recorded during the
   second \textsc{Sunrise} flight\footnote{Since a G-band filter was not part of the \textsc{Sunrise} instrumentation one could use the 300\,nm observations.}
   \citep{Solanki2017} and MHD simulations retrieved from these observations via the newly developed MHD-Assisted Stokes Inversion (MASI) technique \citep{Riethmueller2017}
   because the new technique provides simulations with a magnetic activity level similar to the observations it matches.

   \begin{acknowledgements}
   This work was partially supported by the BK21 plus program through the National Research Foundation (NRF) funded by the Ministry of Education of Korea.
   \end{acknowledgements}



\begin{thebibliography}{}

   \bibitem[Asensio~Ramos(2009)]{AsensioRamos2009} Asensio~Ramos, A. 2009,
      ApJ, 701, 1032
      
   \bibitem[Barthol et al.(2011)]{Barthol2011} Barthol, P., Gandorfer, A., Solanki, S.~K., Sch\"ussler, M., Chares, B., et al. 2011,
      Sol. Phys., 268, 1

   \bibitem[Beck et al.(2007)]{Beck2007} Beck, C., Bellot Rubio, L.~R., Schlichenmaier, R. \& S\"utterlin, P. 2007,
      A\&A, 472, 607

   \bibitem[Beck \& Rezaei(2009)]{Beck2009} Beck, C. \& Rezaei, R. 2009,
      A\&A, 502, 969

   \bibitem[Beckers \& Schr\"oter(1968)]{Beckers1968} Beckers, J.~M., \& Schr\"oter, E.~H. 1968,
      Sol. Phys., 4, 142

   \bibitem[Berger et al.(1995)]{Berger1995} Berger, T.~E., Schrijver, C.~J., Shine, R.~A., Tarbell, T.~D., Title, A.~M., \& Scharmer, G. 1995,
      ApJ, 454, 531

   \bibitem[Berger et al.(2007)]{Berger2007} Berger, T.~E., Rouppe van der Voort, L., \& L\"ofdahl, M. 2007,
      ApJ, 661, 1272

   \bibitem[Berkefeld et al.(2011)]{Berkefeld2011} Berkefeld, T., Schmidt, W., Soltau, D., Bell, A., Doerr, H.~P., et al. 2011,
      Solar Phys., 268, 103

   \bibitem[Bonet et al.(2012)]{Bonet2012} Bonet, J.~A., Cabello, I., \& S\'anchez Almeida, J. 2012,
      A\&A, 539, A6

   \bibitem[Borrero et al.(2015)]{Borrero2015} Borrero, J.~M., Jafarzadeh, S., Sch\"ussler, M., et al. 2015,
      SSRv

   \bibitem[Buehler et al.(2015)]{Buehler2015} Buehler, D., Lagg, A., Solanki, S.~K., \& van~Noort, M. 2015,
      A\&A, 576, A27

   \bibitem[Criscuoli(2013)]{Criscuoli2013} Criscuoli, S. 2013,
      ApJ, 778, 27

   \bibitem[Criscuoli \& Uitenbroek(2014)]{Criscuoli2014} Criscuoli, S. \& Uitenbroek, H. 2014,
      A\&A, 562, L1

   \bibitem[Deinzer et al.(1984)]{Deinzer1984} Deinzer, W., Hensler, G., Sch\"ussler, M., \& Weisshaar, E. 1984,
      A\&A, 139, 435

   \bibitem[Domingo et al.(2009)]{Domingo2009} Domingo, V., Ermolli, I., Fox, P., Fr\"ohlich, C., Haberreiter, M., et al. 2009,
      Space Sci. Rev., 145, 337

   \bibitem[Dunn \& Zirker(1973)]{Dunn1973} Dunn, R.~B. \& Zirker, J.~B. 1973,
      Sol. Phys., 33, 281

   \bibitem[Ermolli et al.(2013)]{Ermolli2013} Ermolli, I., Matthes, K., Dudok de Wit, T., Krivova, N.~A., Tourpali, K., et al. 2013,
      Atmos. Chem. Phys., 13, 3945

   \bibitem[Fr\"ohlich(2013)]{Froehlich2013} Fr\"ohlich, C. 2013,
      Space Sci. Rev., 176, 237

   \bibitem[Frutiger(2000)]{Frutiger2000a} Frutiger, C. 2000,
      Inversion of Zeeman Split Stokes Profiles: Application to solar and stellar surface structures,
      Ph.D. Thesis, Institute of Astronomy, ETH Z\"urich, No.~13896

   \bibitem[Frutiger et al.(2000)]{Frutiger2000b} Frutiger, C., Solanki, S. K., Fligge, M., \& Bruls, J. H. M. J. 2000,
      A\&A, 358, 1109

   \bibitem[Galsgaard \& Nordlund(1996)]{Galsgaard1996} Galsgaard, K. \& Nordlund, \AA{}. 1996,
      J. Geophys. Res., 101, 13445

   \bibitem[Gandorfer et al.(2011)]{Gandorfer2011} Gandorfer, A., Grauf, B., Barthol, P., Riethm\"uller, T.~L., Solanki, S.~K., et al. 2011,
      Sol. Phys., 268, 35

   \bibitem[Gray et al.(2010)]{Gray2010} Gray, L.~J., Beer, J., Geller, M., Haigh, J.~D., Lockwood, M., et al. 2010,
      Rev. Geophys., 48, RG4001

   \bibitem[Haigh et al.(2010)]{Haigh2010} Haigh, J.~D., Winning, A.~R., Toumi, R., \& Harder, J.~W. 2010,
      Nature, 467, 696

   \bibitem[Harder et al.(2009)]{Harder2009} Harder, J.~W., Fontenla, J.~M., Pilewskie, P., Richard, E.~C., \& Woods, T.~N. 2009,
      Geophys. Res. Lett., 36, 7801

   \bibitem[Ishikawa et al.(2007)]{Ishikawa2007} Ishikawa, R., Tsuneta, S., Kitakoshi, Y., Katsukawa, Y., Bonet, J.~A., \& et al. 2007,
      A\&A, 472, 911

   \bibitem[Jafarzadeh et al.(2013)]{Jafarzadeh2013} Jafarzadeh, S., Solanki, S.~K., Feller, A., Lagg, A., Pietarila, A., Danilovic, S., Riethm\"uller, T.~L., \& Mart\'{\i}nez Pillet, V. 2013,
      A\&A, 549, A116

   \bibitem[Ji et al.(2016)]{Ji2016} Ji, Kai-Fan, Xiong, J.-P., Xiang, Y.-Y., Feng, S., Deng, H., \& et al. 2016,
      RAA, 16, 78

   \bibitem[Kahil et al.(2017)]{Kahil2017} Kahil, F., Riethm\"uller, T.~L., \& Solanki, S.~K. 2017,
      ApJ, submitted

   \bibitem[Keys et al.(2011)]{Keys2011} Keys, P.~H., Mathioudakis, M., Jess, D.~B., Shelyag, S., Crockett, P.~J., Christian, D.~J., \& Keenan, F.~P. 2011,
      ApJ, 740, L40

   \bibitem[Keys et al.(2014)]{Keys2014} Keys, P.~H., Mathioudakis, M., Jess, D.~B., Mackay, D.~H., \& Keenan, F.~P. 2014,
      A\&A, 566, A99

   \bibitem[Keys et al.(2013)]{Keys2013} Keys, P.~H., Mathioudakis, M., Jess, D.~B., Shelyag, S., Christian, D.~J., \& Keenan, F.~P. 2013,
      MNRAS, 428, 3220

   \bibitem[Khomenko et al.(2003)]{Khomenko2003} Khomenko, E.~V., Collados, M., Solanki, S.~K., Lagg, A., \& Trujillo~Bueno, J. 2003,
      A\&A, 408, 1015

   \bibitem[Kn\"olker \& Sch\"ussler(1988)]{Knoelker1988} Kn\"olker, M., \& Sch\"ussler, M. 1988,
      A\&A, 202, 275

   \bibitem[Kobel et al.(2011)]{Kobel2011} Kobel, P., Solanki, S.~K., \& Borrero, J.~M. 2011,
      A\&A, 531, A112

   \bibitem[Kobel et al.(2012)]{Kobel2012} Kobel, P., Solanki, S.~K., \& Borrero, J.~M. 2012,
      A\&A, 542, A96

   \bibitem[Krivova et al.(2003)]{Krivova2003} Krivova, N.~A., Solanki, S.~K., Fligge, M., \& Unruh, Y.~C. 2003,
      A\&A, 399, L1

   \bibitem[Krivova et al.(2006)]{Krivova2006} Krivova, N.~A., Solanki, S.~K., \& Floyd, L. 2006,
      A\&A, 452, 631

   \bibitem[Lagg et al.(2016)]{Lagg2016} Lagg, A., Solanki, S.~K., Doerr, H.-P., Mart\'{\i}nez~Gonz\'{a}lez, M.~J., Riethm\"uller, T., et al. 2016,
      A\&A, in press, http://dx.doi.org/10.1051/0004-6361/201628489

   \bibitem[Larkin et al.(2000)]{Larkin2000} Larkin, A., Haigh, J.~D., \& Djavidnia, S. 2000,
      Space Sci. Rev., 94, 199

   \bibitem[Lawrence et al.(1993)]{Lawrence1993} Lawrence, J.~K., Topka, K.~P., \& Jones, H.~P. 1993,
      J. Geophys. Res., 98, 18911

   \bibitem[Lites et al.(2008)]{Lites2008} Lites, B.~W., Kubo, M., Socas-Navarro, H., Berger, T., Frank, Z., \& et al. 2008,
      ApJ, 672, 1237

   \bibitem[London(1994)]{London1994} London, J. 1994,
      Adv. Space Res., 14, 33

   \bibitem[Mart\'{\i}nez~Gonz\'{a}lez et al.(2008)]{MartinezGonzalez2008} Mart\'{\i}nez~Gonz\'{a}lez, M.~J., Asensio~Ramos, A., L\'{o}pez~Ariste, A., \& Manso~Sainz, R. 2008,
      A\&A, 479, 229

   \bibitem[Mart\'{\i}nez Pillet et al.(2011)]{MartinezPillet2011} Mart\'{\i}nez Pillet, V., del Toro Iniesta, J.~C., \'Alvarez-Herrero, A., Domingo, V., Bonet, J.~A., \& et al. 2011,
      Sol. Phys., 268, 57

   \bibitem[Mehltretter(1974)]{Mehltretter1974} Mehltretter, J.~P. 1974,
      Sol. Phys., 38, 43

   \bibitem[Muller(1983)]{Muller1983} Muller, R. 1983,
      Sol. Phys., 85, 113
      
   \bibitem[Muller \& Roudier(1984)]{Muller1984} Muller, R. \& Roudier, Th. 1984,
      Sol. Phys., 94, 33

   \bibitem[Narayan \& Scharmer(2010)]{Narayan2010} Narayan, G. \& Scharmer, G.~B. 2010,
      A\&A, 524, A3
      
   \bibitem[Orozco Su\'{a}rez et al.(2007)]{OrozcoSuarez2007} Orozco Su\'{a}rez, D., Bellot Rubio, L.~R., del~Toro~Iniesta, J.~C., Tsuneta, S., Lites, B.~W., \& et al. 2007,
      ApJ, 670, L61

   \bibitem[Orozco Su\'{a}rez et al.(2010)]{OrozcoSuarez2010} Orozco Su\'{a}rez, D., Bellot Rubio, L.~R., V\"ogler, A., \& del~Toro~Iniesta, J.~C. 2010,
      A\&A, 518, A2

   \bibitem[Rabin(1992)]{Rabin1992} Rabin, D. 1992,
      ApJ, 390, L103

   \bibitem[Riethm\"uller et al.(2017)]{Riethmueller2017} Riethm\"uller, T.~L., Solanki, S.~K., Barthol, P., Gandorfer, A., Gizon, L., \& et al. 2017,
      ApJ, in press

   \bibitem[Riethm\"uller et al.(2014)]{Riethmueller2014} Riethm\"uller, T.~L., Solanki, S.~K., Berdyugina, S.~V., Sch\"ussler, M., Mart\'{\i}nez Pillet, V., Feller, A., Gandorfer, A., \& Hirzberger, J. 2014,
      A\&A, 568, A13
      
   \bibitem[Riethm\"uller et al.(2010)]{Riethmueller2010} Riethm\"uller, T.~L., Solanki, S.~K., Mart\'{\i}nez~Pillet, V., Hirzberger, J., Feller, A., \& et al. 2010,
      ApJ, 723, L169

   \bibitem[R\"ohrbein et al.(2011)]{Roehrbein2011} R\"ohrbein, D., Cameron, R., \& Sch\"ussler, M. 2011,
      A\&A, 532, A140

   \bibitem[Romano et al.(2012)]{Romano2012} Romano, P., Berrilli, F., Criscuoli, S., Del Moro, D., Ermolli, I., \& et al. 2012,
      Sol. Phys., 280, 407

   \bibitem[Rouppe van der Voort et al.(2005)]{Rouppe2005} Rouppe van der Voort, L.~H.~M., Hansteen, V.~H., Carlsson, M., Fossum, A., Marthinussen, E., \& et al. 2005,
      A\&A, 435, 327

   \bibitem[R\"uedi et al.(1992)]{Rueedi1992} R\"uedi, I., Solanki, S.~K., \& Rabin, D. 1992,
      A\&A, 261, L21

   \bibitem[S\'anchez Almeida et al.(2001)]{SanchezAlmeida2001} S\'anchez Almeida, J., Asensio Ramos, A., Trujillo Bueno, J., \& Cernicharo, J. 2001,
      ApJ, 555, 978

   \bibitem[S\'anchez Almeida et al.(2004)]{SanchezAlmeida2004} S\'anchez Almeida, J., M\'arquez, I., Bonet, J.~A., Dom\'{\i}nguez Cerde\~na, I., \& Muller, R. 2004,
      ApJ, 609, L91

   \bibitem[Schnerr \& Spruit(2011)]{Schnerr2011} Schnerr, R.~S. \& Spruit, H.~C. 2011,
      A\&A, 532, A136

   \bibitem[Sch{\"u}ssler et al.(2003)]{Schuessler2003} Sch\"ussler, M., Shelyag, S., Berdyugina, S., V\"ogler, A., \& Solanki, S.~K. 2003,
      ApJ, 597, L173

   \bibitem[Shelyag et al.(2004)]{Shelyag2004} Shelyag, S., Sch\"ussler, M., Solanki, S.~K., Berdyugina, S.~V., \& V\"ogler, A. 2004,
      A\&A, 427, 335

   \bibitem[Solanki(1986)]{Solanki1986} Solanki, S.~K. 1986,
      A\&A, 168, 311

   \bibitem[Solanki(1987)]{Solanki1987} Solanki, S. K. 1987,
      The Photospheric Layers of Solar Magnetic Fluxtubes,
      Ph.D. Thesis, Institute of Astronomy, ETH Z\"urich, No.~8309

   \bibitem[Solanki(1989)]{Solanki1989} Solanki, S.~K. 1989,
      A\&A, 224, 225

   \bibitem[Solanki et al.(2010)]{Solanki2010} Solanki, S.~K., Barthol, P., Danilovic, S., Feller, A., Gandorfer, A., et al. 2010,
      ApJ, 723, L127

   \bibitem[Solanki et al.(2013)]{Solanki2013} Solanki, S.~K., Krivova, N.~A., \& Haigh, J.~D. 2013,
      Ann. Rev. A\&A, 51, 311

   \bibitem[Solanki et al.(2017)]{Solanki2017} Solanki, S.~K., Riethm\"uller, T.~L., Barthol, P., Danilovic, S., Deutsch, W., et al. 2017,
      ApJ, submitted

   \bibitem[Solanki \& Stenflo(1984)]{Solanki1984} Solanki, S.~K. \& Stenflo, J.~O. 1984,
      A\&A, 140, 185

   \bibitem[Spruit(1976)]{Spruit1976} Spruit, H.~C. 1976,
      Sol. Phys., 50, 269  

   \bibitem[Steiner et al.(2001)]{Steiner2001} Steiner, O., Hauschildt, P.~H., \& Bruls, J. 2001,
      A\&A, 372, L13

   \bibitem[Steiner \& Rezaei(2012)]{Steiner2012} Steiner, O. \& Rezaei, R. 2012,
      ASPC, 455, 35

   \bibitem[Stenflo(1973)]{Stenflo1973} Stenflo, J.~O. 1973,
      Solar Phys., 32, 41

   \bibitem[Stenflo(2010)]{Stenflo2010} Stenflo, J.~O. 2010,
      A\&A, 517, A37

   \bibitem[Title \& Berger(1996)]{Title1996} Title, A.~M., \& Berger, T.~E. 1996,
      ApJ, 463, 797

   \bibitem[Topka et al.(1997)]{Topka1997} Topka, K.~P., Tarbell, T.~D., \& Title, A.~M. 1997,
      ApJ, 484, 479

   \bibitem[Tsuneta et al.(2008)]{Tsuneta2008} Tsuneta, S., Ichimoto, K., Katsukawa, Y., Nagata, S., Otsubo, M., \& et al. 2008,
      Sol. Phys., 249, 167

   \bibitem[Utz et al.(2009)]{Utz2009} Utz, D., Hanslmeier, A., M\"ostl, C., Muller, R., Veronig, A., \& Muthsam, H. 2009,
      A\&A, 498, 289

   \bibitem[Utz et al.(2013)]{Utz2013} Utz, D., Jur\v{c}\'ak, J., Hanslmeier, A., Muller, R., Veronig. A., \& K\"uhner, O. 2013,
      A\&A, 554, A65

   \bibitem[V\"ogler(2005)]{Voegler2005a} V\"ogler, A. 2005,
      Mem. Soc. Astr. Ital., 76, 842

   \bibitem[V\"ogler et al.(2005)]{Voegler2005b} V\"ogler, A., Shelyag, S., Sch\"ussler, M., Cattaneo, F., Emonet, T., \& Linde, T. 2005,
      A\&A, 429, 335

   \bibitem[Wiehr et al.(2004)]{Wiehr2004} Wiehr, E., Bovelet, B., \& Hirzberger, J. 2004,
      A\&A, 422, L63

   \bibitem[Willson \& Hudson(1988)]{Willson1988} Willson, R.~C., \& Hudson, H.~S. 1988,
      Nature, 332, 810

   \bibitem[Zakharov et al.(2005)]{Zakharov2005} Zakharov, V., Gandorfer, A., Solanki, S.~K., \& L\"ofdahl, M. 2005,
      A\&A, 437, L43

\end{thebibliography}
\end{document}